\documentclass[reprint,aps,prl,twocolumn,superscriptaddress,showpacs,noeprint]{revtex4-2}
\usepackage{graphicx}
\usepackage[percent]{overpic}
\usepackage{dcolumn}
\usepackage{bm}
\usepackage{color}
\usepackage{amssymb} 
\usepackage{amsmath}
\usepackage{verbatim}
\usepackage{float}
\usepackage{subfig}
\usepackage{lmodern}
\usepackage[utf8]{inputenc}
\usepackage[T1]{fontenc}
\usepackage[english]{babel}

\usepackage{ragged2e}
\DeclareCaptionJustification{justified}{\justifying}
\renewcommand\Re{\operatorname{Re}}
\renewcommand\Im{\operatorname{Im}}

\newcommand{\rt}{\right}
\newcommand{\lf}{\left}

\newcommand{\s}{\text{s}}

\renewcommand{\i}{\text{in}}

\newcommand{\br}{\mathbf r}

\newcommand{\eq}[1]{Eq.~(\ref{#1})}
\newcommand{\Nel}{N_{\text{el}}}

\newcommand{\bG}{\mathbf G}
\newcommand{\mueven}{\mu_{\text{even}}}
\newcommand{\muodd}{\mu_{\text{odd}}}
\newcommand{\bmj}{\boldsymbol{\mathfrak j}}
\newcommand{\qe}{\text{q.e.}}

\newcommand{\commentout}[1]{\ignorespaces}

\begin{document}
\title{Microscopic electron dynamics in nonlinear optical response of solids}
\author{Daria Popova-Gorelova}
\email[]{daria.gorelova@cfel.de}
\affiliation{Center for Free-Electron Laser Science, DESY, Notkestrasse 85, D-22607 Hamburg, Germany}
\affiliation{Department of Physics, Universit\"at Hamburg, Jungiusstrasse 9, D-20355 Hamburg, Germany} 
\affiliation{The Hamburg Centre for Ultrafast Imaging, Universit\"at Hamburg, Luruper Chaussee 149, D-22761 Hamburg, Germany}
\author{Vladislav A.~Guskov}
\affiliation{Center for Free-Electron Laser Science, DESY, Notkestrasse 85, D-22607 Hamburg, Germany}
\affiliation{Moscow Institute of Physics and Technology (State University), Institutskiy Pereulok 9, 141701 Dolgoprudny, Russia}
\author{Robin Santra}
\email[]{robin.santra@cfel.de}
\affiliation{Center for Free-Electron Laser Science, DESY, Notkestrasse 85, D-22607 Hamburg, Germany}
\affiliation{Department of Physics, Universit\"at Hamburg, Jungiusstrasse 9, D-20355 Hamburg, Germany} 
\affiliation{The Hamburg Centre for Ultrafast Imaging, Universit\"at Hamburg, Luruper Chaussee 149, D-22761 Hamburg, Germany}
\date{\today}

\begin{abstract}

We investigate the microscopic properties of the nonlinear optical response of crystalline solids within Floquet theory, and demonstrate that optically-induced microscopic charge distributions display complex spatial structure and nontrivial properties. Their spatial symmetry and temporal behavior are governed by crystal symmetries. We find that even when a macroscopic optical response of a crystal is forbidden, the microscopic optical response can, in fact, be nonzero. In such a case, the optically-induced charge redistribution can be considerable, even though the corresponding Fourier component of the time-dependent dipole moment per unit cell vanishes. We develop a method that makes it possible to completely reconstruct the microscopic optically-induced charge distributions by means of subcycle-resolved x-ray-optical wave mixing. We also show how, within this framework, the direction of the instantaneous microscopic optically-induced electron current flow can be revealed.
 
\end{abstract}
\maketitle

In recent years, laser-driven electron dynamics has gained considerable attention due to remarkable achievements in the field of nonlinear optics including the generation of high-harmonics (HHG) in solids \cite{GhimireNature11, LuuNature15}, optical-field-induced currents in dielectrics \cite{SchiffrinNature12}, manipulation of electric properties of a dielectric with the electric field of light \cite{SchultzeNature12}, control of coherent Bloch oscillations \cite{KuehnPRL10, SchubertNature14}, subcycle terahertz nonlinear optical effects \cite{ChaiPRL18}, coherent control of currents in semiconductors using synthesized optical waveforms \cite{SederbergNatPhot20} and many other intriguing phenomena \cite{SchoetzACSPhotonics19}. These achievements motivated theoretical and experimental studies to understand the mechanisms behind these phenomena \cite{KuehnPRB10, GertsvolfJPhB10, MitrofanovPRL11, KruchininPRB13, HiguchiPRL14, ZhokhovPRL14, VampaNature15, WachterNJPh15, HawkinsPRA15, NdabashimiyeNature16, YouNature16, Tancogne-DejeanPRL17, KruchininRMP18, KlemkeNatComm19,UzanNatPhot20}. At the same time, due to the significant progress of attosecond science, sub-femtosecond x-ray pulses can now be generated \cite{TeichmannNatComm16, HuangPRL17, ParcApplSci18, LiNatComm17, KaertnerNIMPRSA16, DurisNatPhot20}. Such ultrashort x-ray pulses enable real-time measurements of electron dynamics with sub-nanometer spatial resolution. 

In this article, we develop a method that employs ultrafast nonresonant x-ray scattering to probe in real time charge and electron current distributions within a unit cell of a crystal during the interaction with an optical field. We show that spatial symmetry and temporal behavior of microscopic optically-induced charge and electron-current distributions are encoded in the time and momentum dependence of subcycle-resolved x-ray-optical wave-mixing (scr-XROWM) signals. Based on their connection, we develop a method to completely reconstruct microscopic linear and nonlinear charge rearrangements and the direction of electron current flow induced by optical excitation, by means of scr-XROWM.

We describe the interaction of the driving field with a crystal within the framework of the Floquet formalism \cite{ShirleyPR65}. This is a powerful theoretical concept that has been applied to diverse phenomena, such as quantum engineering of novel states of matter aided by a periodic excitation \cite{OkaPRB09, StruckScience11, AidelsburgerPRL11, WangScience13, GoldmanPRX14, SentefNature15, HuebenerNature17, OkaARCMP19,NuskeArxiv20} or nonperturbative processes driven by an intense laser field  \cite{TzoarPRB75, FaisalPRA97, ChuPhRep04, SantraPRA04, ButhPRL07, HiguchiPRL14}. Specifically, we employ the Floquet-Bloch framework, which is capable of a nonperturbative material-specific description of laser-dressed spatially periodic structures \cite{HsuPRB06, FaisalPRA97, TzoarPRB75}. Properties of Floquet-Bloch systems have mainly been analyzed with a focus on HHG \cite{Ben-TalJPhB93, FaisalPRA97, HsuPRB06, HiguchiPRL14, MoiseyevPRA15, IkedaPRA18} or band-structure modification by light \cite{TzoarPRB75, MartinezPRB02, DimitrovskiPRA17, IkedaPRA18}. Here, we apply the Floquet-Bloch formalism as an elegant and insightful framework to calculate laser-induced charge and electron current distributions in real space and real time.

Throughout this paper, we consider a laser-dressed crystal in a state that is characterized by a single Floquet state. As we discuss in Ref.~\onlinecite{Citepaper}, this regime of light-matter interaction includes conventional linear- and nonlinear-optics experiments, in which a single-mode electromagnetic field perturbatively interacts with an optical crystal. It also concerns modern experiments in which a crystal nonperturbatively interacts with an ultrashort light pulse, but wherethe radiation spectrum still consists of harmonic peaks \cite{GhimireNature11, LuuNature15, KuehnPRL10, SchubertNature14, YouNature16}.

In this regime, the electron density of the laser-dressed system evolves in time as \cite{Popova-GorelovaPRB18}
\begin{align}
\rho(\br,t) = \sum_{\mu}e^{i\mu\omega t}\widetilde\rho_\mu(\br)\label{eq_FourierDensity},
\end{align}
where $\omega$ is the frequency of the driving electromagnetic field. The amplitudes $\widetilde\rho_\mu(\br)$ are optically-induced charge distributions that give rise to a $\mu$th-order macroscopic optical polarization $\widetilde{\mathbf P}_{\mu}\propto\int d^3 r \br \widetilde\rho_{\mu}(\br)$. We describe the interaction of a crystal with the electromagnetic field within dipole approximation. All equations are in atomic units. The volume integral of the time-independent part of the electron density gives the number of electrons $\int d^3 r \widetilde\rho_{0}(\br)=\Nel$, reproducing the volume integral of the unperturbed electron density. The volume integral of the density amplitudes of a nonzero order that enter in the time-dependent part of $\rho(\br,t)$ is zero, $\int d^3 r\widetilde \rho_{\mu\neq0}(\br) = 0$. These relations indicate that the interaction of a crystal with light leads to a dynamical redistribution of charges, which has positive and negative regions relative to the field-free electron density. These positive and negative regions coherently oscillate, and cancel each other when volume integrated. The positively charged regions are due to electron holes in valence bands and negatively charged regions are due to electrons in conduction bands.

We consider crystals that obey time-reversal symmetry, but not necessarily obey inversion symmetry. The driving electromagnetic field has an electric field evolving as $\mathbf E_{\text{em}}\sin(\omega t)$ by assumption. Analyzing in Ref.~\onlinecite{Citepaper} how time-reversal symmetry of the electronic Hamiltonian influences the solution of the Floquet-Bloch Hamiltonian, we find that even-order density amplitudes $\widetilde\rho_{\mueven}(\br)$ are real functions, whereas odd-order density amplitudes $\widetilde\rho_{\muodd}(\br)$ are purely imaginary. This property has an important consequence for the time dependence of the electron density, which becomes
\begin{align}
\rho(\br,t) =&\widetilde\rho_0(\br)-\sum_{\muodd\ge 1}\varrho_{\muodd}\sin(\muodd\omega t)\\
&+\sum_{\mueven\ge 2}\varrho_{\mueven}\cos(\mueven\omega t).\nonumber
\end{align}
Here, we redefined the density amplitudes by means of functions
\begin{align}
&\varrho_{\mueven}(\br)=2\Re[\widetilde\rho_{\mueven}(\br)],\quad \varrho_{\muodd}(\br)=2 \Im[\widetilde\rho_{\muodd}(\br)],
\end{align}
which are real for both even and odd $\mu$. Thus, we find that odd-order induced charge distributions oscillate as harmonics in phase with an electric field and even-order induced charge distributions oscillate as harmonics in phase with the vector potential.


We now study how time-reversal symmetry influences the oscillations of the electron current density. Properties of HHG have been investigated using the Floquet-Bloch formalism in several studies \cite{HsuPRB06, HiguchiPRL14, MoiseyevPRA15, IkedaPRA18}. Here, we analyze the electron current density, which is the microscopic property that determines HHG. We obtain that the electron current density evolves in time as
\begin{align}
\mathbf j(\br,t) =& -\sum_{\muodd\ge 1}\bmj_{\muodd}(\br)\cos(\muodd\omega t)\nonumber\\
&-\sum_{\mueven\ge 2}\bmj_{\mueven}(\br)\sin(\mueven\omega t),\label{Eq_osc_current}
\end{align}
where the $\bmj_{\mu}(\br)$ are real-valued amplitudes of the electron current density (see Ref.~\onlinecite{Citepaper} for details). Thus, the electron current density distributions oscillate with a phase shifted by $\pi/2$ with respect to the oscillations of the charge distributions of the same order. It follows from the continuity equation that their amplitudes are connected via the relation $\operatorname{div}\bmj_\mu(\br) = -\mu\omega\varrho_{\mu}(\br)$ .

If a crystal also has inversion symmetry, all even-order density amplitudes are centrosymmetric, whereas all odd-order density amplitudes are antisymmetric (see Ref.~\onlinecite{Citepaper}):
\begin{align}
&\varrho_{\mueven}(\br) = \varrho_{\mueven}(-\br),\quad \varrho_{\muodd}(\br) = -\varrho_{\muodd}(-\br).\label{Eq_den_inv}
\end{align}
Current density amplitudes have opposite symmetry properties in this case
\begin{align}
\bmj_{\mueven}(\br) = -\bmj_{\mueven}(-\br),\quad \bmj_{\muodd}(\br) = \bmj_{\muodd}(-\br).\label{Eq_curr_inv}
\end{align}
Since the volume integral of functions that are antisymmetric is zero, the volume integral of the even-order current density amplitudes is zero, $\int d^3 r\bmj_{\mueven}(\br) = 0$, which leads to the well-known selection rule that even-order harmonics from crystals invariant under inversion symmetry are forbidden \cite{YarivBook}. Thus, time-reversal symmetry of a crystal  governs the temporal behavior of optical response, whereas the spatial symmetry of optically-induced charge and electron-current distributions is determined by the spatial symmetry of a crystal.

\begin{figure}
\centering
\includegraphics[width=0.49\textwidth]{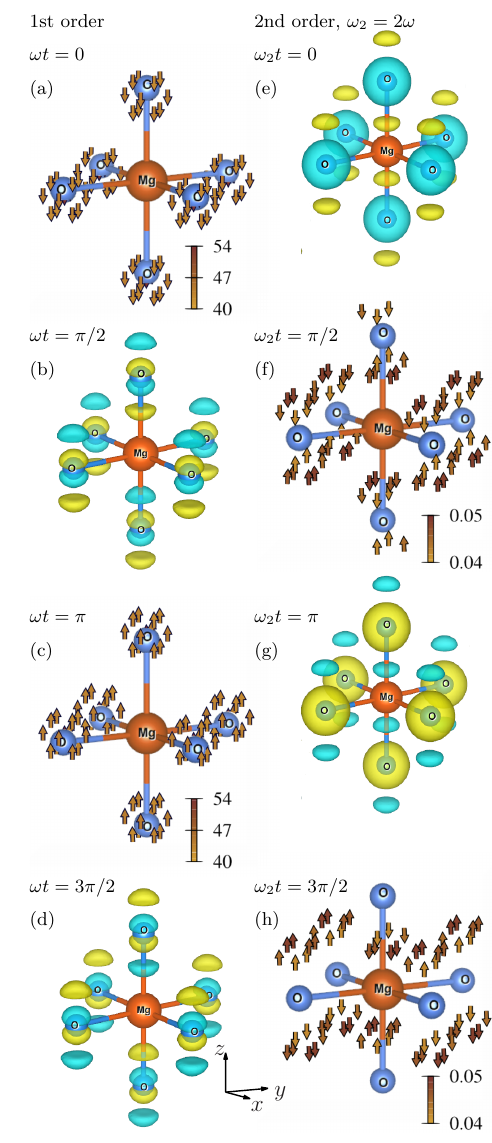}
\caption{The first- and second-order microscopic optical response of a MgO crystal at different phases of the driving electromagnetic field polarized along the $z$ direction. A cut of a unit cell centered around the Mg atom is shown. The first and second columns show the oscillations of the electron density and the electron current density with frequencies $\omega$ and $2\omega$, respectively.   }
\label{MgO_Den_Curr}
\end{figure}



Figure \ref{MgO_Den_Curr} shows the calculated microscopic response of a MgO crystal as a function of the phase of the driving electromagnetic field (see Ref.~\onlinecite{Citepaper} for details). Here, we consider a driving optical field with an intensity of $I_{\text{em}}=2\times 10^{12}$ W/cm$^2$, a photon energy of 1.55 eV, and polarization axis $\boldsymbol\epsilon=(0,0,1)$. We have chosen MgO, since its microscopic optical response turned out to be very illustrative. For the purpose of intuitive visual representation, the electron current densities are plotted on a sparse grid and only vectors with magnitudes $|\bmj_{\mu}(\br)|$ larger than a certain minimum threshold are shown. The magnitudes of $\bmj_{\mu}(\br)$ are color coded and their values are in atomic units. The minimum threshold for $|\bmj_{\mu}(\br)|$ used is the minimum value of the corresponding color box. The electron densities are represented in terms of an isosurface using VESTA \cite{MommaJAC11}. The yellow and blue colors, respectively, represent negative and positive charges relative to the field-free electron density.


The first column in Fig.~\ref{MgO_Den_Curr} displays the first-order oscillations of the electronic state, i.e., the oscillations of the electron density as $-\varrho_{1}\sin(\omega t)$ and the electron current density as $-\bmj_{1}(\br)\cos(\omega t)$ in response to the driving electromagnetic field. MgO has inversion symmetry and first-order electron density amplitudes are antisymmetric with respect to a center of symmetry. The positive and negative charges in Fig.~\ref{MgO_Den_Curr}(b) alternate along the $z$ axis parallel to the optical-field polarization. Clearly, such a charge distribution has a dipole moment along the $z$ direction and is consistent with the first-order polarization of MgO being aligned with the electric field.


The second-order macroscopic polarizability tensor of the MgO crystal is zero because of inversion symmetry. Surprisingly, we find that the second-order {\it microscopic} optical response is nonzero, as shown in the second column of Fig.~\ref{MgO_Den_Curr}. It comprises the oscillations of the electron current density as $-\bmj_{2}(\br)\sin(2\omega t)$ and the oscillations of the electron density as $\varrho_{2}\cos(2\omega t)$, which are both nonzero. The explanation for this phenomenon is that an even-order electron density amplitude of a centrosymmetric crystal is also centrosymmetric. Therefore, the charge distribution in Fig.~\ref{MgO_Den_Curr}(e) has no dipole moment and results in zero {\it macroscopic} polarization. Nonzero second-order microscopic optical response can also be understood by looking at the electron current density in Figs.~\ref{MgO_Den_Curr}(f) and (i). In Ref.~\onlinecite{Citepaper}, we show that the volume integral of $\bmj_{\mu}(\br)$ determines the $\mu$th-order polarization if $\bmj_{\mu}(\br)$ on the boundary of a unit cell fanishes, which is the case as follows from our calculation. In agreement with \eq{Eq_curr_inv}, $\bmj_{2}(\br)$ is antisymmetric and cancels out macroscopically, leading to zero second-order polarization. 

One may think that the second-order microscopic response vanishes, since $|\bmj_{2}|$ is small in comparison to $|\bmj_{1}|$ in Fig.~\ref{MgO_Den_Curr}. In Ref.~\onlinecite{Citepaper}, we performed several calculations to check this.  The third-order harmonic in laser-dressed MgO at similar parameters of the optical field has been observed \cite{YouNature16}.  The calculated $|\bmj_{3}|$ of laser-dressed MgO is even smaller than $|\bmj_{2}|$ and, thus, microscopic second-order response is indeed considerable. In addition, we calculated the microscopic optical response of GaAs, a crystal without inversion symmetry. We compared the second-order response to dressing fields polarized along the $(1,1,1)$ and $(1,0,0)$ directions. In the latter case, the second-order macroscopic optical response is zero according to the macroscopic polarization tensor of GaAs  \cite{YarivBook}. We found that $\varrho_{2}(\br)$ and $\bmj_{2}(\br)$ at different polarizations have comparable magnitudes, but their spatial structures are different. In the case of polarization along $(1,0,0)$, $\bmj_{2}(\br)$ has such symmetry that its volume integral is zero. Thus, even when the $\mu$th-order macroscopic polarization of a crystal vanishes, its $\mu$th-order microscopic optical response can be considerable.

In the following, we develop a method to reconstruct microscopic optical response by means of scr-XROWM. The idea to probe optically-induced charge distributions with an x-ray-optical wave-mixing (XROWM) signal, dates back to the 1970s \cite{FreundPRL70, EisenbergerPRA71}. Since its experimental realization at the x-ray free-electron laser facility Linac Coherent Light Source (LCLS), where the microscopic linear optical response of a diamond crystal was detected \cite{GloverNature12},  XROWM techniques have again moved into the focus of research \cite{SchoriPRL17,RouxelPRL18,CohenPRR19,Popova-GorelovaPRB18}. In Ref.~\onlinecite{Popova-GorelovaPRB18}, we developed a general theoretical framework to describe the interaction of general Floquet systems with an x-ray pulse and applied it to describe a subcycle-unresolved measurement. As we will show, the method presented here, which is based on subcycle-resolved measurement, provides much more deeper insights into laser-driven electron dynamics.

\begin{figure}
\centering
\includegraphics[width=0.49\textwidth]{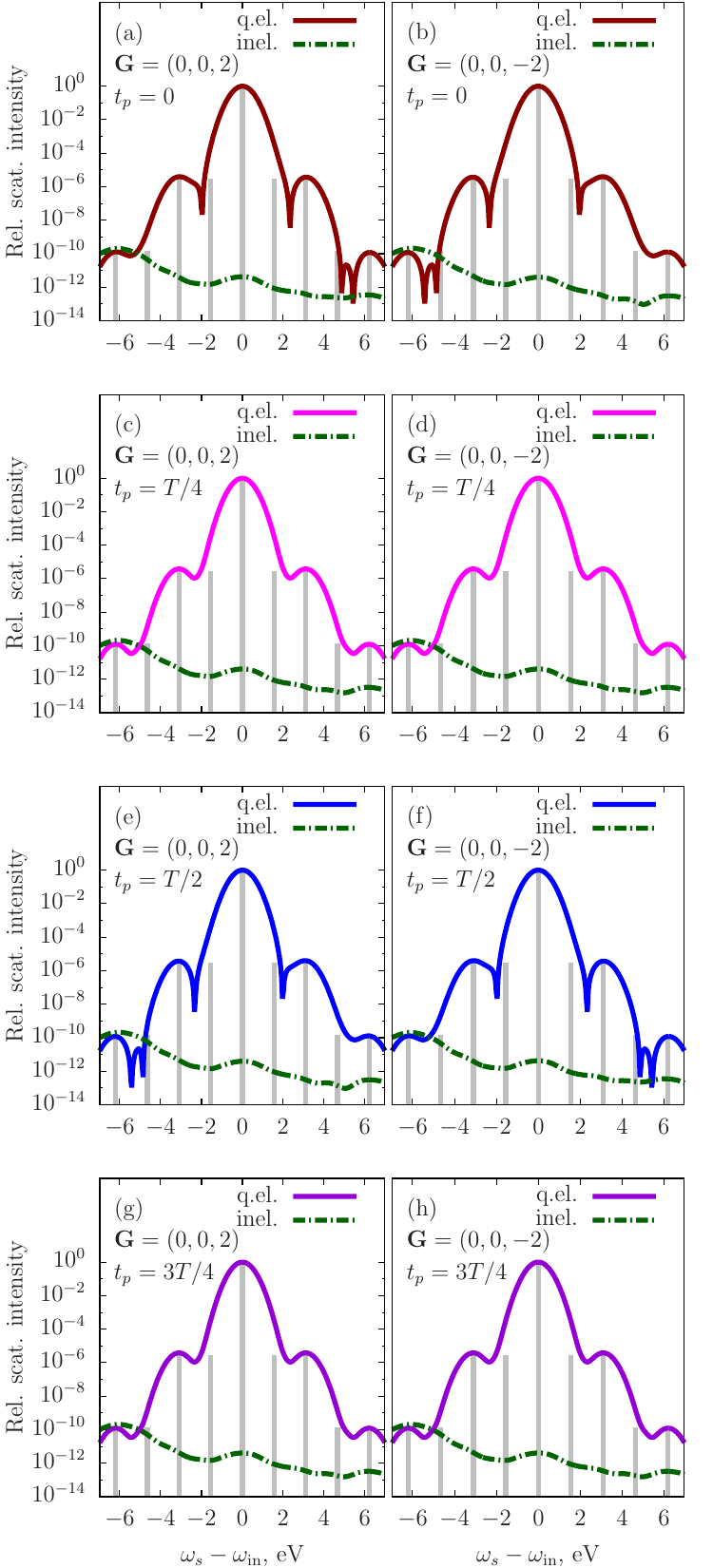}
\caption{Intensities of quasielastic and inelastic x-ray scattering signals at $\mathbf G = (0,0,2)$ and $\mathbf G = (0,0,-2)$ from a laser-dressed MgO crystal at different probe-pulse arrival times as a function of $\omega_\s-\omega_\i$. The intensities are normalized to the intensity of the main Bragg peak of MgO at $\mathbf G = (0,0,2)$. The gray vertical lines are situated at the positions of the side peaks, $\mu\omega$, and their heights correspond to their relative intensities.}
\label{Fig_Braggs}
\end{figure}

We consider an ultrashort nonresonant x-ray pulse with photon energy $\omega_\i$ and Gaussian-shaped electric field for probing the electronic state of a crystal during its interaction with the optical field. Figure \ref{Fig_Braggs} shows calculated x-ray scattering signals from laser-dressed MgO at reciprocal lattice vectors $\bG = (0,0,2)$ and $\bG = (0,0,-2)$ at different probe-pulse arrival times. The scattering signals are normalized to the main Bragg peak at $\bG = (0,0,2)$, which is centered at the scattered energy $\omega_\s = \omega_\i$. The scattering signal is the sum of quasielastic and inelastic contributions, $P_{\text{tot.}} = P_{\text{q.e.}} + P_{\text{inel.}}$ \cite{Popova-GorelovaPRB18}. The quasielastic contribution results in an XROWM signal at reciprocal lattice vectors $\bG$, i.e., the appearance of the $\mu$th-order side peaks to the Bragg peaks of a crystal at scattered energies $\omega_s=\omega_\i+\mu\omega$. With decreasing duration of the x-ray pulse, the spectral bandwidth of the side peaks becomes broader and they start to interfere. Here, we have chosen the probe-pulse duration such that only nearest-neighbor side peaks interfere. Namely, the x-ray pulse duration is 2 fs for the considered optical field with an optical period of $T=2.67$ fs. In this case, $P_{\text{q.e.}}$ can be represented as  (see Ref.~\onlinecite{Citepaper} for details) 
\begin{align}
&P_{\qe} (\bG) = \sum_{\mu}\mathcal S_\mu\lf|\mathcal P_{\mu}(\bG)\rt|^2+\sum_{\mu}P_{\mu\leftrightarrow\mu+1} (\bG,t_p)\mathcal S'_{\mu+1/2}\label{Eq_Pqe_Side_Intefer}.
\end{align}
Here, $\mathcal P_{\mu}(\bG)=|\mathcal P_{\mu}|e^{i\alpha_\mu(\bG)}=\int d^3r e^{i\bG\cdot\br}\varrho_{\mu}(\br)$ is a Fourier component of an optically-induced charge distribution. $P_{\mu\leftrightarrow\mu+1} $ depends on $\mathcal P_{\mu}$ and $\mathcal P_{\mu+1}$ as will be shown below. $\mathcal S_\mu$ and $\mathcal S'_{\mu+1/2}$ are Gaussian-shaped functions of $\omega_\s$ (see Ref.~\onlinecite{Citepaper}), $t_p$ is the time of x-ray-pulse arrival relative to a reference time $t=0$, when $\omega t=0$. The first term in \eq{Eq_Pqe_Side_Intefer} represents Gaussian-shaped side peaks centered at $\omega_s=\omega_\i+\mu\omega$ with time-independent amplitudes.  The second term represents the interference between nearest-neighbor side peaks that are Gaussian-shaped peaks centered at $\omega_s=\omega_\i+(\mu+1/2)\omega$. As we will show next, these terms provide most interesting insights into electron dynamics.

Scr-XROWM in Figs.~\ref{Fig_Braggs}(a), (b), (e) and (f) is not centrosymmetric with respect to $\bG$. Since the time-independent part of $P_{\qe}$ is centrosymmetric, this effect is due to time-dependent interference terms. We analyze the consequence of time-reversal symmetry for their time- and momentum-dependence in Ref.~\onlinecite{Citepaper}. We find that the interference terms have an antisymmetric contribution that oscillates in phase with the first-order electron current density 
\begin{align}
P_{\mu\leftrightarrow\mu+1} (\bG,t_p) -P_{\mu\leftrightarrow\mu+1} (-\bG,t_p)\propto\cos(\omega t_p),
\end{align}
and a centrosymmetric contribution that, if present, oscillates in phase with the first-order charge distribution
\begin{align}
P_{\mu\leftrightarrow\mu+1} (\bG,t_p) + P_{\mu\leftrightarrow\mu+1} (-\bG,t_p)\propto\sin(\omega t_p).
\end{align}
More generally, if the temporal resolution allows resolving interference terms between $\mu$th- and $(\mu+\Delta\mu)$th-order side peaks, their antisymmetric and centrosymmetric parts would also oscillate in phase with the $\Delta\mu$th-order electron current density and electron density, respectively. Thus, the symmetry breaking in scr-XROWM signals in reciprocal space is a direct signature of microscopic laser-driven electron currents in real space.  

The interference terms are determined by
\begin{align}
P_{\mu\leftrightarrow\mu+1} (\bG,t_p)= &(-1)^{\mu+1}\lf|\mathcal P_{\mu}(\bG)\rt| \lf|\mathcal P_{\mu+1}(\bG)\rt| \nonumber\\
& \times \sin\lf[\omega t_p-\alpha_\mu(\bG)+\alpha_{\mu+1}(\bG)\rt].\label{Eq_interfer_term}
\end{align}
This means that the time dependence of scr-XROWM depends on the phase of the Fourier components of optically-induced charge distributions. For example, the signal from the laser-dressed MgO in Fig.~\ref{Fig_Braggs} oscillates as $\cos(\omega t_p)$. The unperturbed density of MgO is centrosymmetric, i.e., $\alpha_{0}(\bG)=n\pi$, where $n$ is either 0 or 1 depending on $\bG$. According to \eq{Eq_interfer_term}, this means that the phases of optically-induced charge distributions are $\alpha_{\muodd}=\pm\pi/2$ and $\alpha_{\mueven}=n\pi$. Thus, even-order charge distributions are antisymmetric, and odd-order charge distribution are symmetric for MgO in agreement with \eq{Eq_den_inv}. We analyze the difference of signals at opposite $\bG$ shown in Fig.~\ref{Fig_Braggs} in spectral regions around $\omega_\s-\omega_\i=0.5\,\omega$ and $\omega_\s-\omega_\i=1.5\,\omega$, and obtain that $\mathcal P_{1} (\bG)/\mathcal P_{0} (\bG)=-1.7\times 10^{-3}i$ and $\mathcal P_{2} (\bG)/\mathcal P_{0} (\bG)=-2.0\times10^{-3}$ for $\bG=(0,0,2)$ (see Ref.~\onlinecite{Citepaper}). The scr-XROWM signal from laser-dressed GaAs has contributions oscillating both as $\sin(\omega t_p)$ and as $\cos(\omega t_p)$ (see Ref.~\onlinecite{Citepaper}). In this case, we reconstruct the phases of the charge distribution by looking at both the difference and the sum of the signals at opposite $\bG$. With data at various $\bG$, optically-induced microscopic optical response can be completely reconstructed from the delay dependence of scr-XROWM.

Since the charge distributions and the electron current densities are connected via the equation of continuity, their Fourier transforms are connected via
\begin{align}
\int d^3 r e^{i\mathbf G\cdot\mathbf r} \varrho_\mu(\br)= -\frac1{\mu\omega} \bG\cdot \int d^3 r e^{i\mathbf G\cdot\mathbf r} \bmj_\mu(\br).\label{Eq_Four_curr}
\end{align}
The first-order electron current density in Fig.~\ref{MgO_Den_Curr} is parallel to $\bG=(0,0,2)$. The signals at opposite $\bG$ in Figs.~\ref{Fig_Braggs}(a) and (b) switch with the signals in Figs.~\ref{Fig_Braggs}(f) and (e) when the direction of electron current changes its sign [cf.~Figs.~\ref{MgO_Den_Curr}(a) and (c)]. Thus, not only the time dependence, but also the direction of microscopic electron currents is encoded in scr-XROWM.


The widely-used concept that optically-induced charge separation merely gives rise to a dipole moment fails on the atomic scale. Induced charge distributions have a rich structure and various symmetry features. Even when the induction of a macroscopic polarization is forbidden, charges still rearrange within a unit cell and electron currents are formed. Such microscopic optical response can be detected with XROWM, and can be even larger than the first-order optical response for some scattering directions.  If XROWM signals are subcycle-resolved, they encode the phases of the Fourier components of charge distributions, the time dependence of microscopic electron currents and their direction. An atomically-resolved view into light-matter interactions will provide a deeper understanding of optically-driven electron dynamics and prompt further developments of nonlinear optics towards technological applications.

\section*{Acknowledgment}
We acknowledge valuable discussions with David A.~Reis and Matthias Fuchs. Daria Popova-Gorelova acknowledges funding from the Volkswagen Foundation through a Freigeist Fellowship. V.~A.~Guskov acknowledges support from the DESY Summer Student Program.


\end{document}